 \documentclass[final,5p,times,twocolumn,authoryear]{elsarticle}

\usepackage{amssymb}
\usepackage{lipsum}
\usepackage{amsmath}
\usepackage{listings}
\usepackage[dvipsnames]{xcolor}
\usepackage[colorlinks = true, linkcolor = blue, urlcolor  = blue, citecolor = blue, anchorcolor = blue]{hyperref}
\usepackage[T1]{fontenc}
\usepackage[scaled=0.825]{FiraMono}
\newcommand{\MYhref}[3][blue]{\href{#2}{\color{#1}{#3}}}

\begin{document}

\begin{frontmatter}



\title{Multi-Dimensional Wasserstein Distance Implementation in Scipy}


\author[first]{Zehao Lu}
\affiliation[first]{organization={Utrecht University},
            addressline={Heidelberglaan 8}, 
            city={Utrecht},
            postcode={3584 CS}, 
            state={Utrecht},
            country={Netherlands}}

\begin{abstract}
The Wasserstein distance, also known as the Earth mover distance or optimal transport distance, is a widely used measure of similarity between probability distributions. This paper presents an linear programming based implementation of the multi-dimensional Wasserstein distance function in Scipy, a powerful scientific computing package in Python. Building upon the existing one-dimensional \texttt{scipy.stats.wasserstein\_distance} function, our work extends its capabilities to handle multi-dimensional distributions. 
To compute the multi-dimensional Wasserstein distance, we developed an implementation that transforms the problem into a linear programming problem. We utilized the scipy linear programming solver to effectively solve this transformed problem.
The proposed implementation includes thorough documentation and comprehensive test cases to ensure accuracy and reliability. The resulting feature is set to be merged into the main Scipy development branch and will be included in the upcoming release, further enhancing the capabilities of Scipy in the field of multi-dimensional statistical analysis.
\end{abstract}




\end{frontmatter}




\section{Introduction}
\label{introduction}

The Wasserstein distance, known by alternative names such as the Earth mover distance or optimal transport distance, serves as a measure of similarity between two probability distributions (\cite{vaserstein1969markov, OLKIN1982257}). In the discrete case, the Wasserstein distance represents the cost associated with the optimal transport plan required to move from one set of samples (or distribution) to another (\cite{kantorovich1960mathematical}).
Following its initial introduction in the Monge problem, extensive research has been dedicated to studying the Wasserstein distance over many years (\cite{bogachev2012monge}). 
It has been widely used in many areas to compare discrete distributions. For example, it was used to compare color histograms in computer vision, measuring the document distance, distribution distance in econometric models, or as a similarity metric for anomaly detection (\cite{rubner2000earth, wan2005earth, 10.2307/j.ctt1q1xs9h.4, pereira2019learning}).
Within the context of the WGAN neural network framework, it has been employed as a loss function (\cite{arjovsky2017wasserstein}).

Given two probability mass functions, $u$ and $v$, the first Wasserstein distance between the distributions is:

$$
l_1 (u, v) = \inf_{\pi \in \Gamma (u, v)} \int_{\mathbb{R} \times
        \mathbb{R}} |x-y| \mathrm{d} \pi (x, y)
$$
where $\Gamma (u, v)$ is the set of (probability) distributions on $\mathbb{R} \times \mathbb{R}$ whose marginals are $u$ and $v$ on the first and second factors respectively.

In the case where both inputs come from one-dimensional distributions, the Wasserstein distance is equivalent to the energy distance, and calculating the energy distance is a straightforward process (\cite{ramdas2017wasserstein}). However, this equivalence does not hold true in a multi-dimensional metric space. Consequently, computing the numerical solution for the multi-dimensional Wasserstein distance between given samples is considerably more challenging compared to the one-dimensional case.
In this report, I summarize my contribution to implementing the multi-dimensional wasserstein distance function as well as corresponding documentation and tests in \MYhref[PineGreen]{https://scipy.org/}{Scipy}, which is the most comprehensive and powerful scientific computing package in Python (\cite{2020SciPy-NMeth}).
The new feature is under the same namespace of the existing one-dimensional \texttt{scipy.stats.wasserstein\_distance} function.
The current documentation of the \texttt{wasserstein\_distance} function can be found in the \MYhref[PineGreen]{https://docs.scipy.org/doc/scipy/reference/generated/scipy.stats.wasserstein_distance.html\#scipy.stats.wasserstein_distance}{released document}. 
My work on multi-dimensional Wasserstein distance is going to be merged into the Scipy main developing branch after another round of review and will be released in the upcoming version.

\section{Method}
To begin with, I will provide a concise overview of the Monge problem expressed in discrete form.
The problem at hand is precisely the focus and objective of the \texttt{scipy.stats.wasserstein\_distance}
function, aiming to tackle and resolve it.

Let the finite point sets $\{x_i\}$ and $\{y_j\}$ denote the support set of probability mass function $u$ and $v$ respectively.
As state in the previous section, the Wasserstein distance between $u$ and $v$ is,

$$
        l_1 (u, v) = \inf_{\pi \in \Gamma (u, v)} \int_{\mathbb{R} \times
        \mathbb{R}} |x-y| \mathrm{d} \pi (x, y)
$$

Let $D$ denote the distance matrix $[d_{ij}]$ in which $d_{ij}$ is the distance from $x_i$ to $y_j$, and $\Gamma$ denotes matrix $[\gamma_{ij}]$ in which $\gamma_{ij}$ is a positive value representing the amount of probability mass transported from $u(x_i)$ to $v(y_i)$. Therefore the matrix $\Gamma$ is a well-defined transport plan if and only if summing over the rows of $\Gamma$ should give the source distribution $u$: $\sum_j \gamma_{ij} = u(x_i)$ holds for all $i$ and summing over the columns of $\Gamma$ should give the target distribution $v$: $\sum_i \gamma_{ij} = v(y_j)$ holds for all $j$. And there is

$$
        l_1(u, v) = min(\sum_i\sum_j D \circ \Gamma | \\ \forall i: \sum_j \gamma_{ij} = u(x_i), \forall j: \sum_i \gamma_{ij} = v(y_j))
$$

This report exclusively focuses on the computation of the Wasserstein distance between discrete and finite samples or distributions on a discrete and finite support set.

\subsection{One Dimensional Case}

\begin{figure}[htb]
    \label{fig0}
	\centering 
	\includegraphics[width=0.4\textwidth]{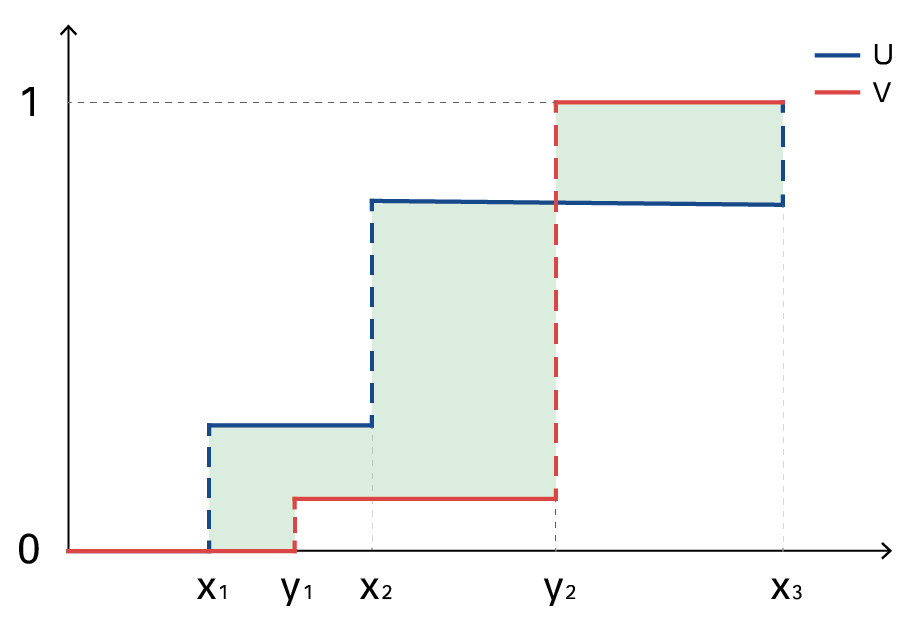}	
	\caption{Example of the CDF distance. The blue and red step function  represent the CDF curve of distribution $U$ and $V$, which support are $\{x_1, x_2, x_3\}$ and $\{y_1, y_2\}$ correspondingly.}
    \label{fig1}
\end{figure}
In the 1-dimensional case, let $U$ and $V$ denote the respective CDFs of $u$ and $v$, the Wasserstein distance also equals to the first-order CDF distance, according to \cite{ramdas2017wasserstein, bellemare2017cramer}:

\begin{equation}\label{eq1}
    l_1(u, v) = \int_{-\infty}^{+\infty} |U-V| d(x, y)
\end{equation}

Figure \ref{fig1} provides a comprehensive visualization of the 1-CDF distance, represented by the green shaded area between the CDF curves. In the discrete case, computing the 1-CDF distance becomes straightforward by summing the product of the differences between the input samples and the differences in CDFs.

To show that the 1-CDF distance in equivalent to the Wasserstein-1 distance in the distance case (the proof of continious case can be found in \cite{}), we provide a brief illustration showing that the Wasserstein-1 distance between the input samples ($U$, $V$) is always equal to the area between the CDF curves of $U$ and $V$. It is important to note that the following explanation is not a formal proof, as the precise mathematical analysis is not the primary focus of this internship project.
\begin{figure}[t]
	\centering 
	\includegraphics[width=0.4\textwidth]{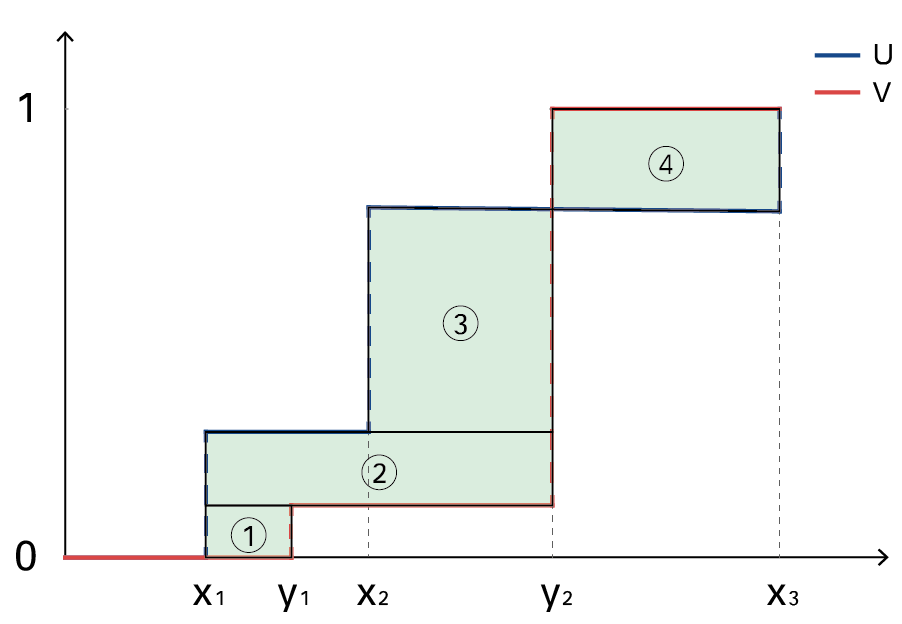}	
	\caption{Example of the CDF distance. Each segmented rectangle area with number are corresponding to a transport move in Figure \ref{fig3}.}
     \label{fig2}
\end{figure}

\begin{figure}[ht]
	\centering
	\includegraphics[width=0.4\textwidth]{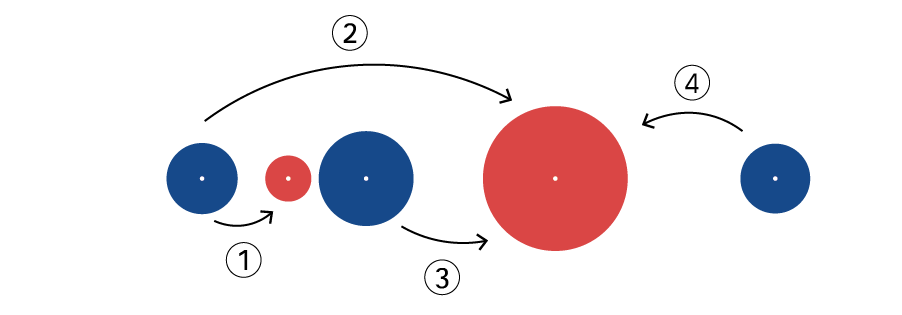}	
	\caption{Example of a optimal transport plan T. Each colored circle denotes a probability distribution at the circle's center, the area of the circle (approximately) shows the probability mass. Each arrow represents a transport from $U$ to $V$ and corresponds to a rectangle area in Figure \ref{fig2}.}
     \label{fig3}
\end{figure}
Suppose we have two distributions $U$ and $V$ and their discrete support set $\{x_i\}$ and $\{y_j\}$, where $x_i$ and $y_j$ are values on $\mathbb{R}$. Their CDF curves given in figure \ref{fig1}, it is trivial that the area of green area between the two CDF is equal to the value of the 1-CDF distance, as defined in formula \ref{eq1}. Hence, the remaining task is to demonstrate that the Wasserstein distance is equal to the green area. 
To accomplish this, we establish that an optimal transport plan involves sorting the input samples (or the union of discrete support) $\{x_i\} \cup \{y_j\}$ and iteratively assigning the probability mass from the smallest position to its nearest 'available' target position based on probability mass. This process is exemplified in Figure \ref{fig2} and Figure \ref{fig3}.

Firstly, it is evident that the depicted transport plan $T$ satisfies the condition where each transport from the source distribution to the target ensures that the maximum available probability mass is transported to "fill the hole" at the target position. Next, we make a contradictory assumption that the illustrated transport plan is not optimal, implying the existence of another transport plan $\tilde{T}$ that achieves greater savings in transport cost. 
Consider the sequence $\{t_1, t_2, ..., t_5\}$ to represent the independent transports in the transport plan illustrated in Figure \ref{fig3}. Similarly, let the sequence ${\tilde{t_i}}$ represent the transports in $\tilde{T}$, sorted based on their source positions. In the case where two or more transports share the same source position, they are further sorted based on their respective target positions.

Let $\tilde{t_k}$ represent the initial transport move in ${\tilde{t_i}}$ that does not exist in $T$. We obtain its source position $\tilde{s_k}$, target position $\tilde{r_k}$, transported probability mass $\tilde{p_k}$, and define the function $\tilde{P_k}$ as follows.

\begin{equation}
       \tilde{P_k}(x) = 
        \begin{cases}
            \tilde{p_k} & \text{if $x \in \{\tilde{s_k}, \tilde{r_k} \} $} \\
            0 & \text{otherwise}
        \end{cases}
\end{equation}

Furthermore, let $t_k$ represent the transport move in $T$ that corresponds to the same order as $\tilde{t_k}$, and its source, target, and transported probability are denoted in the same manner. It should be noted that $k$ is less than or equal to 5; otherwise, $\tilde{T}$ would be identical to $T$. Also, because the $\tilde{t_k}$ is the first transport move in $\tilde{T}$ that does not exist in $T$, there is $s_k = \tilde{s_k}$ and $r_k < \tilde{r_k}$.
We have, 
\begin{equation}
\begin{split}
    t_k &: s_k \rightarrow r_k, P_k(s_k) = U(s_k) - \sum_{i<k} P_i(s_k)\\
    \tilde{t_k} &: \tilde{s_k} \rightarrow \tilde{r_k}, \tilde{P_k}(s_k) < P_k(s_k)
\end{split}
\end{equation}

 Note that $\tilde{P_k}(s_k) < P_k(s_k)$ is because the illustrated optimal transport $T$ greedily move the available probability mass from the source distribution, so the probability mass transported by $\tilde{t_k}$ must be smaller than those transported by $t_k$. Consider $P_k(s_k) = U(s_k) - \sum_{i<k} P_i(s_k)$ and $P_k(s_k) < V(r_k) - \sum_{i<k} P_i(s_k)$, there is 
\begin{equation}
\begin{split}
   \tilde{P_k}(s_k) & < U(s_k) - \sum_{i<k} P_i(s_k)\\
   \tilde{P_k}(r_k) \leq \tilde{P_k}(s_k) & < V(r_k) - \sum_{i<k} P_i(r_k)
\end{split}
\end{equation}

 which means if we only consider the latter part of transport plan $\tilde{T}^* = \{\tilde{t_i}\}_{i > k}$, there is available probability mass in position $s_k$ and unfilled target position $r_k$.
 
 Based on the above observation, we can deduce the existence of a pair of transport moves in $\tilde{T}$. The first move, $\tilde{t}_{k+1}$, originates from $s_k$, while the second move, $\tilde{t_l}$, is directed towards $r_k$. There is 
\begin{equation}
\begin{split}
    \tilde{t}_{k+1} &: s_k \rightarrow \tilde{r}_{k+1}, \tilde{p}_{k+1}\\
    \tilde{t_l} &: \tilde{s_l} \rightarrow r_k, \tilde{p_l}
\end{split}
\end{equation}
with
\begin{equation}
\tilde{s_l} \geq \tilde{s}_{k+1} = \tilde{s}_{k} = s_k, \tilde{r}_{k+1} > \tilde{r}_{k} > r_k, l \geq k+1
\end{equation}
Therefore one can easily find a better transport plan than $\Tilde{T}$ by slightly adjust $\tilde{t}_{k+1}$ and $\tilde{t}_{l}$ using a small constant $m$.
\begin{equation}
\begin{split}
    \tilde{t}_{k+1} &: s_k \rightarrow \tilde{r}_{k+1}, \tilde{p}_{k+1} - m\\
    \tilde{t_l} &: \tilde{s_l} \rightarrow r_k, \tilde{p_l} + m \\
    m & < min(\tilde{p}_{k+1}, \tilde{p}_{l})
\end{split}
\end{equation}
as the overall cost of the new plan is decreased by $m * (|s_k - \tilde{r}_{k+1}| - |\tilde{s_l} - r_k|)$, which is always positive under the condition that $s_k \leq \tilde{s_l}$ and $\tilde{r}_{k+1} > r_k$. By contradiction, I proved that the optimal transport plan is as illustrated in Figure \ref{fig3} and can be found by greedy weight assignment algorithm. Then it is trivial that the Wasserstein distance is equals to the area between the source and the target distribution, as each of the numbered rectangle area in Figure \ref{fig2} is equals to the cost its corresponding transport move in Figure \ref{fig3}. For example, the area of the number 2 rectangle in Figure \ref{fig2} is $(U(x_1) - V(y_1)) \times |y_2 - x_1|$, and the cost of the number 2 transport move is also $(U(x_1) - V(y_1)) \times |y_2 - x_1|$.

\subsection{Multi Dimensional Case}\footnotemark
\footnotetext{The content in this sub-section is largely based on Vincent Hermann's \MYhref[PineGreen]{https://vincentherrmann.github.io/blog/wasserstein/}{blog} "Wasserstein GAN and the Kantorovich-Rubinstein Duality".}
In the more general (higher dimensional) and discrete case, the solution using the 1-CDF distance from the previous section doesn't hold anymore. Therefore, we present a new solution based on the linear-programming approach.

In practice, our utilization of the linear programming method relies on the internal linear solver from scipy, known as 'highs' \cite{huangfu2015parallelizing}.
This solver is designed to adaptively select between 'highs-ipm' and 'highs-ds' methods based on the properties of the input. 'highs-ipm' is a C++ wrapper for the high-performance an \textbf{interior-point} algorithm, while 'highs-ds' is a C++ wrapper for the HSOL implementation of the high-performance dual revised \textbf{simplex} method.

Let $\Gamma$ denote the transport plan, $D$ denote the
    distance matrix, $u$, $v$ denote the weight or the probability mass and,

\begin{equation} \label{eq2}
\begin{split}
        x &= \text{vec}(\Gamma)          \\
        c &= \text{vec}(D)               \\
        b &= \begin{bmatrix}
                u\\
                v\\
            \end{bmatrix}
\end{split}
\end{equation}

The $\text{vec}()$ function denotes the Vectorization function
that transforms a matrix into a column vector by vertically stacking
the columns of the matrix. 

Same as it is stated previously, the tranport plan $\Gamma$ is a matrix $[\gamma_{ij}]$ in which $\gamma_{ij}$ is a positive value representing the amount of
probability mass transported from $u(x_i)$ to $v(y_i)$.
Summing over the rows of $\Gamma$ should give the source distribution
$u$ : $\sum_j \gamma_{ij} = u(x_i)$ holds for all $i$
and summing over the columns of $\Gamma$ should give the target
distribution $v$: $\sum_i \gamma_{ij} = v(y_j)$ holds for all
$j$.
The distance matrix $D$ is a matrix $[d_{ij}]$, in which
    $d_{ij} = d(x_i, y_j)$.

    Given $\Gamma$, $D$, $b$, the Monge problem can be
    tranformed into a linear programming problem by
    taking $A x = b$ as constraints and $z = c^T x$ as minimization
    target (sum of costs) , where matrix $A$ has the form

\begin{equation} \label{eq3}
    \begin{bmatrix}
        \begin{array} {rrrr|rrrr|r|rrrr}
            1 & 1 & \dots & 1 & 0 & 0 & \dots & 0 & \dots & 0 & 0 & \dots &
                0 \cr
            0 & 0 & \dots & 0 & 1 & 1 & \dots & 1 & \dots & 0 & 0 &\dots &
                0 \cr
            \vdots & \vdots & \ddots & \vdots & \vdots & \vdots & \ddots
                & \vdots & \vdots & \vdots & \vdots & \ddots & \vdots  \cr
            0 & 0 & \dots & 0 & 0 & 0 & \dots & 0 & \dots & 1 & 1 & \dots &
                1 \cr \hline

            1 & 0 & \dots & 0 & 1 & 0 & \dots & \dots & \dots & 1 & 0 & \dots &
                0 \cr
            0 & 1 & \dots & 0 & 0 & 1 & \dots & \dots & \dots & 0 & 1 & \dots &
                0 \cr
            \vdots & \vdots & \ddots & \vdots & \vdots & \vdots & \ddots &
                \vdots & \vdots & \vdots & \vdots & \ddots & \vdots \cr
            0 & 0 & \dots & 1 & 0 & 0 & \dots & 1 & \dots & 0 & 0 & \dots & 1
        \end{array}
    \end{bmatrix}
\end{equation}

The production of $A$ and the transport plan representation vector $x$ is constraint by the formula $Ax = b$. 
The $i$th upper section in $A$ simply suggests that the sum of the $i$th row in $\Gamma$ should be equals to the $i$th weight or probability mass in $u$. 
Similarly, the $i$th lower section in $A$ is suggesting that the sum of the $i$th column in $\Gamma$ should be equals to the $i$th weight or probability mass in $v$.

By solving the dual form of the above linear programming problem (with
solution $y^*$), the Wasserstein distance $l_1 (u, v)$ can
be computed as $b^T y^*$.
To conclude this section, I give the primal and dual forms of the Monge problem.

\begin{equation}\label{eq4}
    \begin{array}{c|c}

\mathbf{primal \ form:} & \mathbf{dual \ form:}\\

\begin{array}{rrcl}
\mathrm{minimize} \ & z & = & \ c^T x, \\
\mathrm{so \ that} \ & A x & = & \ b \\
\mathrm{and}\  & x & \geq &\ 0
\end{array} &

\begin{array}{rrcl}
\mathrm{maximize} \ & \tilde{z} & = & \ b^T y, \\
\mathrm{so \ that} \ & A^T y & \leq & \ c \\ \\
\end{array}

\end{array}
    \end{equation}

\section{Impelementation}
This section provides the implementation details of the proposed function along with a concise explanation of the algorithm. It is followed by a set of unit tests aimed at verifying the accuracy of the proposed function and ensuring its expected behavior. Lastly, we provide a statistical analysis of the algorithm's computational efficiency.

\definecolor{backcolour}{HTML}{F0F0F0}
\definecolor{codegray}{HTML}{E74646}
\definecolor{codepurple}{HTML}{159895}
\definecolor{codegreen}{HTML}{607EAA}
\definecolor{codeorange}{HTML}{FF6D60}

\lstdefinestyle{mystyle}{
    backgroundcolor=\color{backcolour},   
    commentstyle=\color{codegreen},
    keywordstyle=\color{codegray},
    numberstyle=\color{codeorange},
    stringstyle=\color{codepurple},
    basicstyle=\ttfamily\footnotesize,
    breakatwhitespace=false,
    breaklines=true,                 
    captionpos=b,                    
    keepspaces=true,                 
    numbers=left,                    
    numbersep=5pt,                  
    showspaces=false,                
    showstringspaces=false,
    showtabs=true,                  
    tabsize=2
}

\lstset{style=mystyle}

\subsection{Program}

I give the full capacity python code for the Wasserstein distance computation here, the support functions that are call inside the \texttt{wasserstein\_distance} function will be explained later. The doc string in the function is removed as the algorithm is already introduced in the previous section.

The function takes four inputs in total, two positional parameters \texttt{u\_values}, \texttt{v\_values} and two optional parameters \texttt{u\_weights}, \texttt{v\_weights}.
All inputs are array-like objects, the \texttt{u\_values} and \texttt{v\_values} are either 1d or 2d arrays, each of them represents a sample from a probability distribution or the support (set of all possible values) of a probability distribution. 
Note that a \textbf{2d} array actually represents a set of \textbf{multi-dimensional} vectors.
For \texttt{u\_values} and \texttt{v\_values}, each element along the first axis is an observation or possible value. If inputs values are two-dimensional, the second axis represents the dimensionality of the distribution; i.e., each row is a vector observation or possible value. The optional array-like inputs \texttt{u\_weights}, \texttt{v\_weights} represent weights or counts corresponding with the sample or probability masses corresponding with the support values. The sum of elements in \texttt{u\_weights} or \texttt{v\_weights} must be positive and finite. If they are unspecified, each value is assigned the same weight.

\begin{lstlisting}[language=Python]
def wasserstein_distance(u_values, v_values, u_weights=None, v_weights=None):
    r"""
    ...
    """
    m, n = len(u_values), len(v_values)
    u_values = asarray(u_values)
    v_values = asarray(v_values)

    if u_values.ndim > 2 or v_values.ndim > 2:
        raise ValueError('Invalid input values. The inputs must have either one or two dimensions.')

    # if dimensions are not equal throw error
    if u_values.ndim != v_values.ndim:
        raise ValueError('Invalid input values. Dimensions of inputs must be equal.')

    # if data is 1D then call the cdf_distance function
    if u_values.ndim == 1 and v_values.ndim == 1:
        return _cdf_distance(1, u_values, v_values, u_weights, v_weights)

    u_values, u_weights = _validate_distribution(u_values, u_weights)
    v_values, v_weights = _validate_distribution(v_values, v_weights)

    # if number of columns is not equal throw error
    if u_values.shape[1] != v_values.shape[1]:
        raise ValueError('Invalid input values. If two-dimensional, `u_values` and `v_values` must have the same number of columns.')

    # if data contains np.inf then return inf or nan
    if np.any(np.isinf(u_values)) ^ np.any(np.isinf(v_values)):
        return np.inf
    elif np.any(np.isinf(u_values)) and np.any(np.isinf(v_values)):
        return np.nan

    # create constraints
    A_upper_part = sparse.block_diag((np.ones((1, n)), ) * m)
    A_lower_part = sparse.hstack((sparse.eye(n), ) * m)
    # sparse constraint matrix of size (m + n)*(m * n)
    A = sparse.vstack((A_upper_part, A_lower_part))
    A = sparse.coo_array(A)

    # get cost matrix
    D = distance_matrix(u_values, v_values, p=2)
    cost = D.ravel()

    # create the minimization target
    p_u = np.full(m, 1/m) if u_weights is None else u_weights/np.sum(u_weights)
    p_v = np.full(n, 1/n) if v_weights is None else v_weights/np.sum(v_weights)
    b = np.concatenate((p_u, p_v), axis=0)

    # solving LP
    constraints = LinearConstraint(A=A.T, ub=cost)
    opt_res = milp(c=-b, constraints=constraints, bounds=(-np.inf, np.inf))
    return -opt_res.fun
\end{lstlisting}

Let's provide a brief overview of the functions implemented in the above program, organized by lines.
\begin{description}
   \item[Line 1 - line 4] Defining the function, input arguments and give doc string.
   \item[Line 5 - line 7] Measuring the length of the input arrays, calling the \texttt{asarray} function to convert the inputs to \MYhref[PineGreen]{https://scipy.org/}{\texttt{numpy.array}} object.
   \item[Line 9 - line 15] Give error and terminate the program if the input shape are not expected (more than two dimensional or the number of dimension are not equal).
   
   \item[Line 17 - line 18] Call the \texttt{\_cdf\_distance} function if the inputs are 1d. The answer are computed using CDF distance as it is shown in previous section.
   
   \item[Line 20 - line 21] Calling the \texttt{\_validate\_distribution} function to make sure that each of the inputs has the same length as the corresponding weight, all weights are non- negative and the sum of weights are positive and finite.
   
   \item[Line 23 - line 25] Throw error if the input distributions have different dimensionality.
   
   \item[Line 27 - line 31] If the data contains infinite or missing value, return infinity or \texttt{numpy.nan}.
   
   \item[Line 33 - line 38] Separately create the upper and lower part of the constraint matrix $A$ and stack them together, as shown in formula \ref{eq3}. 

   \item[Line 40 - line 42] Compute the distance matrix $D$ and flatten it.

   \item[Line 44 - line 47] If the weights are not specified, create uniform weights and concatenate the weights to get the  minimization target $b$.

   \item[Line 49 - line 52] Solve the dual form of the linear programming problem with constraints and optimization target and return answer.

\end{description}

\subsection{Examples}
Some examples (these examples are also included in the documentation) of the input arguments and the function's output are presented in this section.

\textbf{Example 1}: Compute the Wasserstein distance between one-dimensional inputs. These examples already exists before my commits in this project and they can be found in the \MYhref[PineGreen]{https://docs.scipy.org/doc/scipy/reference/generated/scipy.stats.wasserstein_distance.html\#scipy.stats.wasserstein_distance}{released document} of the \texttt{wasserstein\_distance} function.
\begin{lstlisting}[language=Python]
from scipy.stats import wasserstein_distance
wasserstein_distance([0, 1, 3], [5, 6, 8])
# answer: 5.0
wasserstein_distance([0, 1], [0, 1], [3, 1], [2, 2])
# answer: 0.25
wasserstein_distance([3.4, 3.9, 7.5, 7.8], 
                     [4.5, 1.4],
                     [1.4, 0.9, 3.1, 7.2], 
                     [3.2, 3.5])
# answer: 4.0781331438047861
\end{lstlisting}

\textbf{Example 2}: Compute the Wasserstein distance between two three-dimensional samples, each with two observations.
\begin{lstlisting}[language=Python]
wasserstein_distance([[0, 2, 3], 
                      [1, 2, 5]], 
                     [[3, 2, 3], 
                      [4, 2, 5]])
# answer: 3.0
\end{lstlisting}

\textbf{Example 3}: Compute the Wasserstein distance between two two-dimensional distributions with three and two weighted observations, respectively.

\begin{lstlisting}[language=Python]
wasserstein_distance([[0, 2.75], 
                      [2, 209.3], 
                      [0, 0]],
                     [[0.2, 0.322], 
                      [4.5, 25.1808]],
                     [0.4, 5.2, 0.114], 
                     [0.8, 1.5])
# answer: 174.15840245217169
\end{lstlisting}

\subsection{Tests}

I also added several test in the special class, \texttt{TestWassersteinDistance} , which is designed for test the \texttt{wasserstein\_distance}'s behaviour, as its name suggested. The \texttt{TestWassersteinDistance} class is under the namespace \texttt{scipy.tests}. I list all of the tests and their purpose below in order to provide a convincing result, the existing tests before my commits are tagged with a *.
Note that the tests added in this project are mostly following the property-based fashion using the python unit test package \MYhref[PineGreen]{https://docs.pytest.org/en/7.3.x/}{pytest}, that is, generate random value use for testing, while the existing tests are value-based, in other words, it test the behaviour of the function using hard code value as inputs (\cite{pytestx.y}).

For the sake of readability, I do not give the source code in this section, please click this \MYhref[PineGreen]{https://github.com/scipy/scipy/blob/e0f307133bd5955e4c72dbfd204b468adac13c82/scipy/stats/tests/test_stats.py\#L7316-L7544}{link} to find the original code.

\begin{description}
    \item[\texttt{test\_simple}*]  Test the function for basic distributions, the value of the Wasserstein distance is straightforward.

    \item[\texttt{test\_published\_values}] Compare the result from proposed function against published values and manually computed results. The values and computed result are posted at James D. McCaffrey's \MYhref[PineGreen]{https://jamesmccaffrey.wordpress.com/2018/03/05/earth-mover-distance-wasserstein-metric-example-calculation/}{blog}. 

    \item[\texttt{test\_same\_distribution}*] Any distribution moved to itself should have a Wasserstein distance of zero.

    \item[\texttt{test\_same\_distribution\_nD}] Property-based. Multi-dimensional version for the above test.

    \item[\texttt{test\_shift}*] If the whole distribution is shifted by vector $x$, then the Wasserstein distance should be the norm of $x$.

    \item[\texttt{test\_combine\_weights}*] Assigning a weight $w$ to a value is equivalent to including that value $w$ times in the value array with weight of 1.

    \item[\texttt{test\_collapse}*] Collapsing a distribution to a point distribution at zero is equivalent to taking the average of the absolute values of the values.

    \item[\texttt{test\_collapse\_nD}] Property-based. Collapsing a n-D distribution to a point distribution at zero is equivalent to taking the average of the norm of data.

    \item[\texttt{test\_zero\_weight}*] Values with zero weight have no impact on the Wasserstein distance.

    \item[\texttt{test\_zero\_weight\_nD}] Property-based. Multi-dimensional version for the above test.

    \item[\texttt{test\_inf\_values}*] Infite values can lead to an infinite distance or trigger a RuntimeWarning (and return NaN) if the distance is undefined. I included some mulit-dimensional tests under this method.

    \item[\texttt{test\_multi\_dim\_nD}] Property-based. Adding dimension on distributions do not affect the result.

    \item[\texttt{test\_orthogonal\_nD}] Property-based. Orthogonal transformations do not affect the result of the Wasserstein distance.

    \item[\texttt{test\_error\_code}] Verify whether the raised error code matches the expected value.
\end{description}


%

%

%

%

%
    
%

%
    
%

%
    
%

\subsection{Performance Testing}
This section includes a performance test of the Wasserstein distance function. The test procedure is as follows:
First, I invoke the Wasserstein function on various random data pairs multiple times and measure the computation time. The inputs are adjusted in quantity, ranging from $2^0$ to $2^9$. Both inputs have a fixed dimension of 2, as the dimension only affects the computation time of the distance matrix and is not the main focus of this project.

\begin{figure}[htb]
    \label{fig4}
	\centering 
	\includegraphics[width=0.5\textwidth]{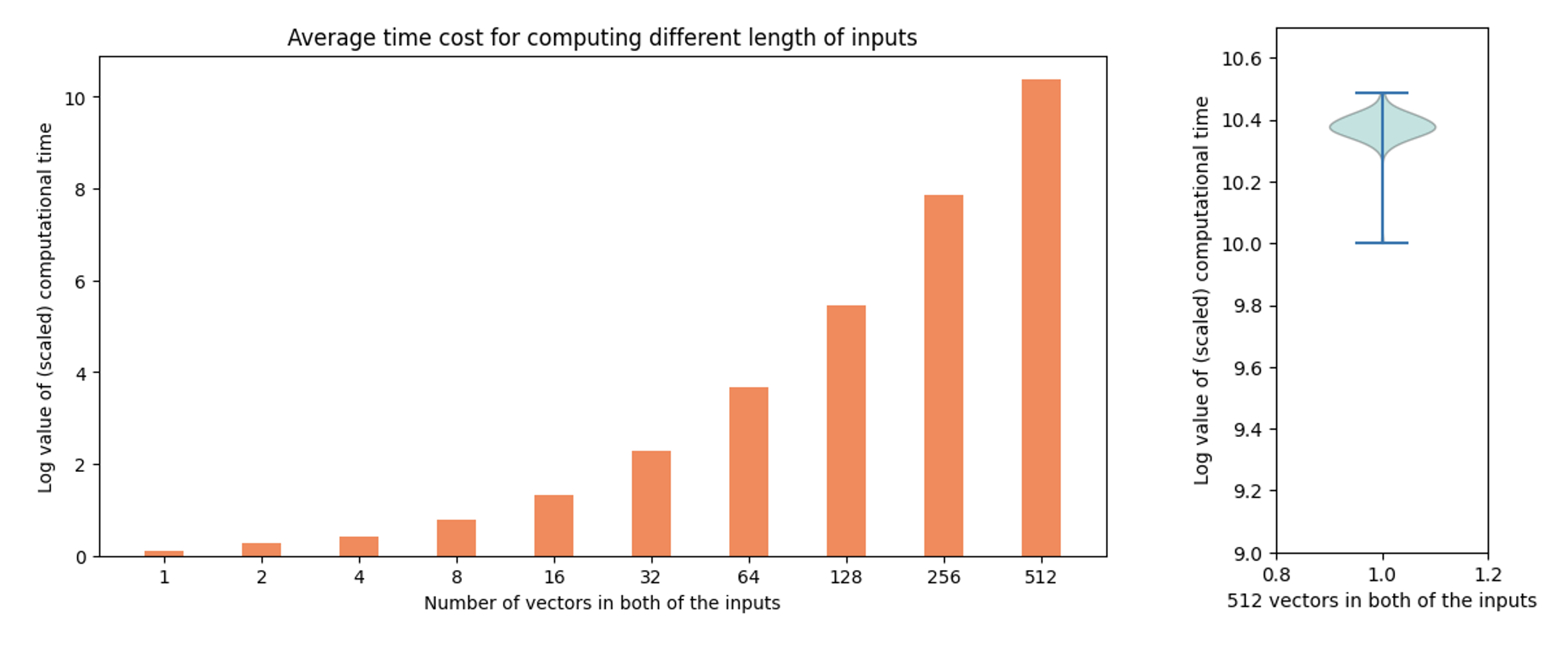}	
	\caption{The computational time of the wasserstein\_distance function when applied to inputs of various sizes. The left plot presents a bar graph depicting the scaled average computational time across different input shapes, which are indicated along the x-axis.
    Meanwhile, the right plot features a violin plot showcasing the distribution of the scaled computational time specifically for inputs with shapes (512, 512).}
\end{figure}

Next, I document the computational time required for various input pairs. For each data quantity, I perform 100 tests and record the corresponding time. To scale the computational time appropriately, I employ the following transformation and record the scaled times:
$$
\hat{t} = log(t-1)
$$
The scaled average time for each data quantity is presented in Figure \ref{fig4} (left), while the distributions are displayed in Figure \ref{fig4} (right). Notably, the majority of the recorded scaled computational times fall within the range of 10.0 to 10.6.

\section{Discussion \& Conclusion}
In this study, we have presented the implementation of the multi-dimensional Wasserstein distance function in Scipy, using the linear programming approach. I have successfully achieved the research objectives, and I have also noted that there is room for improvement in the computational efficiency of the proposed solution. Further research can be conducted to explore the application of different algorithms for calculating the Wasserstein distance, such as the network simplex algorithm or the Sinkhorn algorithm, in order to enhance the efficiency of the computations (\cite{orlin1997polynomial, pham2020unbalanced}).

\appendix

\section{Open source contribution done right}
I will present a step-by-step life cycle of contributing to an open-source software (or packages):
\begin{description}
    \item[familiarize with Github] 
    \MYhref[PineGreen]{https://github.com}{Github} is the biggest and most popular web-based platform and service that provides hosting for software development projects using the Git version control system. 
    It offers a collaborative environment for developers to work on projects, track changes, manage code repositories, and facilitate team collaboration.
    Most of the influential open-source project are hosted on github or have an image hosted.
    \item[Pick a project] Understand the project's goals, features, and existing codebase. Read the documentation, explore the issue tracker, and review any contribution guidelines or coding standards provided. For instance, you can access the \MYhref[PineGreen]{https://docs.scipy.org/doc/scipy/dev/core-dev/index.html}{developer guide for Scipy} directly on their website, which will offer detailed insights and instructions for contributing to the project.
    \item[Select a task] Look for issues or tasks suitable for your skills and interests. For any open source project, if the codebase is hosted on github, there is an issue list. If there is any issue that is attractive to you, you could directly comment and express your willingness to tackle this issue. If there is no such issue, you can also create one you own. In this project, I posted a new issue \MYhref[PineGreen]{https://github.com/scipy/scipy/issues/17290}{here} to summarize the existing debate and requests on the multi-dimensional Wasserstein distance in Scipy.
    \item[Discuss the plans] Engage with the project's community, either through mailing lists, forums, or chat channels. Once the issue attracted some attentions, you can share your intentions and seek guidance to ensure your proposed contribution aligns with the project's vision and to avoid duplicating efforts.
    \item[Set up development environment] Install the necessary dependencies, set up the project locally, and ensure you can build and test the software effectively. For Scipy, \MYhref[PineGreen]{https://docs.conda.io/en/latest/}{Conda} is required for this.
    \item[Create a branch] To contribute to the project, you should begin by forking the repository and creating a new branch dedicated to your contributions. This isolated branch is where you can make your desired changes. \MYhref[PineGreen]{https://git-scm.com/}{Git} is an essential tool for this process, as it is widely recognized and extensively used as a version control system in the field of computer science (\cite{chacon2014pro}). If you are new to Git and would like to get started quickly, you can learn the basics by following this resource: \MYhref[PineGreen]{https://www.atlassian.com/git/tutorials/learn-git-with-bitbucket-cloud}{Learn Git with Bitbucket Cloud}.
    \item[Submit a pull request] Push your branch to your forked repository and submit a pull request (PR) to the original project's repository. Provide a clear and concise description of your changes, explaining their purpose and any relevant details. You can find my pull request for this contribution at \MYhref[PineGreen]{https://github.com/scipy/scipy/pull/17473}{here}.
    \item[Iterate and address feedback] Collaborate with the project maintainers and address any feedback or code review comments promptly. This may take several months or weeks, for this project, it takes 6 months already. Once your changes are viewed by all reviewers, all changes will be wrapped up and merged in the main branch.

\end{description}

\bibliographystyle{elsarticle-harv} 
\bibliography{example}






\end{document}